# High $J_c$ and low anisotropy of hydrogen doped NdFeAsO superconducting thin film

Kazumasa Iida[1,6], Jens Hänisch[2], Keisuke Kondo[1], Mingyu Chen[1], Takafumi Hatano[1,6], Chao Wang[3], Hikaru Saito[4,6], Satoshi Hata[3,5,6], and Hiroshi Ikuta[1]

[1]*Department of Materials Physics, Nagoya University, Chikusa-ku, Nagoya 464-8603, Japan*
[2]*Institute for Technical Physics, Karlsruhe Institute of Technology, Hermann-von-Helmholtz-Platz 1, 76344 Eggenstein-Leopoldshafen, Germany*
[3]*The Ultramicroscopy Research Center, Kyushu University, Nishi-ku, Fukuoka 819-0395, Japan*
[4]*Institute for Materials Chemistry and Engineering, Kyushu University, Kasuga, Fukuoka 816-8580, Japan*
[5]*Faculty of Engineering Sciences, Kyushu University, Kasuga, Fukuoka 816-8580, Japan*
[6]*JST CREST, Kawaguchi, Saitama 332-0012, Japan*

**The recent realisations of hydrogen doped *Ln*FeAsO (*Ln*=Nd and Sm) superconducting epitaxial thin films call for further investigation of their structural and electrical transport properties. Here, we report on the microstructure of a NdFeAs(O,H) epitaxial thin film and its temperature, field, and orientation dependencies of the resistivity and the critical current density $J_c$. The superconducting transition temperature $T_c$ is comparable to NdFeAs(O,F). Transmission electron microscopy investigation supported that hydrogen is homogenously substituted for oxygen. A high self-field $J_c$ of over 10 MA/cm² was recorded at 5 K, which is likely to be caused by a short London penetration depth. The anisotropic Ginzburg-Landau scaling for the angle dependence of $J_c$ yielded temperature-dependent scaling parameters $\gamma_J$ that decreased from 1.6 at 30 K to 1.3 at 5 K. This is opposite to the behaviour of NdFeAs(O,F). Additionally, $\gamma_J$ of NdFeAs(O,H) is smaller than that of NdFeAs(O,F). Our results indicate that heavily electron doping by means of hydrogen substitution for oxygen in *Ln*FeAsO is highly beneficial for achieving high $J_c$ with low anisotropy without compromising $T_c$, which is favourable for high-field magnet applications.**

The Fe-based superconductors (FBS), the second class of high-temperature superconductors beside the cuprates, are considered as possible candidates for high-field magnet applications [1-5]. Among them, *Ln*FeAs(O,F) (*Ln*: Nd and Sm) has the highest depairing current density $J_d$ of ~170 MA/cm² at zero kelvin [6]. Additionally, *Ln*FeAs(O,F) shows the highest superconducting transition temperature $T_c$. These two features together with their high upper critical fields make *Ln*FeAs(O,F) attractive, although the electromagnetic anisotropy is slightly higher than that of other FBS.

Very similar to the partial substitution of fluorine for oxygen in $Ln$FeAsO, hydrogen also leads to electron doping ($O^{2-} \rightarrow H^- + e^-$) [7], resulting in a $T_c$ of up to ~55 K. The distinct difference between H- and F-doping is the substitution limit: $x \leq 0.8$ for $Ln$FeAsO$_{1-x}$H$_x$ [8] in contrast to $x \leq 0.2$ for $Ln$FeAsO$_{1-x}$F$_x$ [9]. Furthermore, a high $T_c$ of ~50 K is maintained in the range $0.13<x<0.43$ for $Ln$FeAsO$_{1-x}$H$_x$ [7]. The growth of $Ln$FeAs(O,H) opens new opportunities to explore how heavily electron doping influences the superconducting properties. However, most of the studies have been carried out on polycrystals [7, 8] or tiny single crystals [10], on which measurements of the transport critical current density are rather complicated. The successful growth of $Ln$FeAs(O,H) epitaxial thin films gives a great opportunity to explore the intrinsic physical properties by electrical transport measurements especially for critical current characteristics, since thin films are the ideal platform for such investigations.

SmFeAs(O,H) epitaxial thin films have recently been grown on single-crystal MgO(001) by a combination of pulsed laser deposition and topotactic chemical reaction through post-annealing with $Ae$H$_2$ ($Ae$=Ca, Sr, Ba, and Mg) powders that serve as hydrogen source [11, 12]. By referring to this hydrogen doping method, we have fabricated H-doped NdFeAsO epitaxial thin films [13]. In this article, we present the electrical transport properties of a NdFeAs(O,H) epitaxial thin film with a thickness of ~24 nm. The film was characterised over a wide temperature range and in magnetic fields up to 14 T.

**Results**

**Microstructure**

Microstructural analysis by transmission electron microscopy (TEM) confirmed that our NdFeAs(O,H) film is almost free of defects in the matrix as well as at the interface (fig. 1a). The atomic-resolution annular dark-field (ADF) image agrees well with the crystal structure of NdFeAs(O,H) projected along the *b*-axis, as shown in the inset of fig. 1a (top left). This ADF image also revealed the atomic arrangement at the NdFeAs(O,H)/MgO interface. The first atomic layer in the NdFeAs(O,H) film exhibits brighter contrast than surroundings, indicating that a Nd layer is firstly formed on the MgO substrate at the beginning of film growth. In this interfacial Nd layer, a large density of dislocations is introduced, as shown in the inset of fig. 1a (bottom left). Those misfit dislocations compensate the large lattice parameter difference, *i.e.*, *a* (NdFeAsO) = 3.99 Å while *a* (MgO) = 4.23 Å, resulting in the defect-free matrix inside the NdFeAs(O,H) film.

Figures 1b and c show magnified ADF images of NdFeAsO and NdFeAs(O,H), respectively, clearly indicating a shrinkage of the lattice in the *c*-axis direction by H substitution for oxygen. The *c*-axis lattice parameter decreased from 8.64 Å to 8.50 Å, as shown in the extracted intensity profiles (fig. 1d). It is reported that the *c*-axis lattice parameter decreases with increasing hydrogen content $x$ in $Ln$FeAsO$_{1-x}$H$_x$ [7, 10, 11] with a rate of $\Delta c/\Delta x \sim -2$-$3 \times 10^{-3}$ Å/at.%. The lattice parameter $c$ of our NdFeAs(O,H) film determined by x-ray diffraction (XRD) was 8.437±0.003 Å, which also supports the lattice shrinkage due to hydrogen doping although the value was slightly shorter than the average value evaluated from TEM. In order to check the homogeneity of hydrogen doping, the $c$-axis lattice parameters in the vicinity of the MgO substrate and near the film surface were compared, resulting in the same value (fig. 1d). This result implies a homogeneous H substitution for oxygen, which guarantees that the transport properties shown below are not affected by local inhomogeneity.

**Resistivity measurements for determining the magnetic phase diagram**
Figures 2a and b summarise the field dependence of resistivity for both major field directions, $H$ parallel to the $ab$-plane and to the $c$-axis. $T_c$ is recorded at 44 K, which is 2 K lower than the as-processed NdFeAs(O,H) film (Supplementary information fig. S1). The reason for the reduced $T_c$ may be that the sample was slightly damaged during bridge fabrication.

A clear shift of $T_c$ to lower temperatures with magnetic fields is observed for both directions. This shift together with a broadening of the transition is more obvious for $H \parallel c$ than $\parallel ab$. The temperature dependencies of $H_{c2}$, fig. 2c, show slopes of -11.8 T/K for $H \parallel ab$ and -2.7 T/K for $H \parallel c$ in the range $0 \leq \mu_0 H \leq 4$ T. Hence, the anisotropy of $H_{c2}$ near $T_c$ is around $\gamma_{Hc2}$=4.4, which is lower than for NdFeAs(O,F) film ($\gamma_{Hc2}$=5.1) of similar thickness (22 nm) [14]. For cuprate superconductors, it has been shown that the anisotropy decreased with doping because of the increase in the interlayer coupling [15]. The decreased $\gamma_{Hc2}$ for NdFeAs(O,H) may be explained similarly.

The temperature dependence of the irreversibility field $H_{irr}$, fig. 2c, for $H \parallel ab$ shows a kink around 4 T, which is due to a matching field effect. This effect has the same origin as reported for the 22-nm thick NdFeAs(O,F) film in ref. [14]. The matching field corresponds to the film thickness and is related to the Bean-Livingston barrier [16]. Hence, the origin of this matching field effect differs distinctly from the one commonly observed for $H \parallel c$ in $RE$Ba$_2$Cu$_3$O$_7$ ($RE$BCO, $RE$: rare earth elements) films containing highly correlated columnar defects with diameter of a few nano meters [17-18]. The results of $H_{irr}$ for $H \parallel c$ are discussed later.

**Pinning potential**
The field dependence of the activation energy, $U_0(H)$, for vortex motion can be estimated from linear fits to the Arrhenius plots of $\rho(T)$, figs. 3a and b, under the assumption of $U(T,H)=U_0(H)(1-T/T_c)$ leading to $\ln\rho(T,H)=\ln\rho_0(H)-U_0(H)/T$ and $\ln\rho_0(H)=\ln\rho_{0f}+U_0(H)/T_c$ [19]. Here, $\rho_{0f}$ is a pre-factor. For both main orientations and all fields, $U_0(H)$ is systematically larger than for the 22 nm-thick NdFeAs(O,F) film reported earlier [14], e.g., for $H \parallel c$ at 1 T, 4.2×10$^3$ K for NdFeAs(O,H) and $3.5 \times 10^3$ K for NdFeAs(O,F). $U_0(H)$ shows a power law relation $H^{-\alpha}$ for both main orientations, fig. 3c, except for $H \parallel c$ in high fields, where $U_0(H)$ is better expressed by $U_0(H) \sim H^{-0.5}(1-H/H^*)^2$ ($\mu_0 H^* \sim 48$ T). This fitting formula has been used for polycrystalline $MgB_2$ samples by Thompson *et al*. who argued that the exponents should be the same as the ones in the pinning force density analysis [20]. These exponents (i.e., 0.5 and 2) suggest that Kramer's scaling for the pinning force density holds, which will be discussed later. For both directions, the exponent α is 0.07 at low fields, which can be explained by single vortex pinning [21]. The distinct feature for $H \parallel ab$ is that α changes from 0.07 to ~1 in the range 2~4 T, followed by 0.34 above 4 T, although the value of α~1 may contain somewhat large uncertainty as we have only three data points in this field regime. Nevertheless, the exponent α=1 indicates that collective pinning is dominating in this field regime [21]. The transition field at which the exponent α changes from 1 to 0.34 corresponds to the matching field shown in fig. 2c. It is intriguing that the pinning mechanism for $H \parallel ab$ changes from single vortex pinning to collective pinning, followed by plastic pinning (*i.e.*, α~0.5 [22]).

**Field dependence of $J_c$ and the pinning force density**

Field dependence of $J_c$ for both $H \parallel ab$ and $\parallel c$, and the corresponding pinning force density $F_p$ are summarised in figs. 4 a-d. Self-field $J_c$ of NdFeAs(O,H) at 5 K exceeds 10 MA/cm$^2$. Another film with a $T_c$ of 45 K prepared by the same condition showed even a self-field $J_c$ of over 17 MA/cm$^2$ at 4 K [13]. These values are higher than our best-performing NdFeAs(O,F) film of similar thickness (22 nm) [14] (purple line in Fig. 4a-d), albeit the reduced temperature ($t=T/T_c \sim 0.114$) of NdFeAs(O,H) was higher than that of NdFeAs(O,F) ($t \sim 0.093$). Below 20 K, $J_c$ is rather insensitive against the applied field for $H \parallel ab$ (fig. 4a) and $F_p$ shows a linear increase above 4 T, indicative of strong single-vortex pinning. The reason for that is intrinsic pinning and will be discussed later. The elemental pinning force density per length for intrinsic pinning can be calculated by $f_p' = \frac{1}{\mu_0}\frac{dF_p}{dH}\phi_0$. The respective $f_p$' are $8.0 \times 10^{-5}$ N/m at 5 K, $4.2 \times 10^{-5}$ N/m at 10 K, $1.5 \times 10^{-5}$ N/m at 15 K, and $1.7 \times 10^{-6}$ N/m at 20 K. On the other hand, for $H \parallel c$, $J_c$ monotonously decreases with increasing applied field, which reflects the absence of macroscopic defects in our film (*i.e.*, a clean microstructure as can be seen in fig. 1).

In order to understand the pinning mechanism for $H \parallel c$, the normalised pinning force densities $f_p=F_p/F_{p,max}$ were plotted as a function of the reduced field $h=H/H_{irr}$. $H_{irr}$ was evaluated from $J_c$-$H$ characteristics with a criterion of 1.4 kA/cm$^2$ in the temperature range $20 \leq T \leq 35$ K. The fit of $f_p \sim h^p(1-h)^q$ to each $f_p$ at given temperatures is shown in Supplementary information fig. S2, and the resulting fitting parameters *p* and *q* are plotted as a function of temperature (fig. 4e). Although both *p* and *q* show a slight temperature dependence, the respective values of *p* and *q* are almost close to 0.5 and 2, suggesting that the Kramer model for shear breaking of the flux line lattice is mainly responsible for depinning [23].

For $T \leq 15$ K $H_{irr}$ cannot be evaluated from $J_c$-$H$ characteristics due to the experimental limitation. Hence, $H_{irr}$ was determined from fits to the pinning force density, on the assumption that the Kramer model prevails in the whole $T$ range [i.e., $(p, q)$=(0.5, 2)].

The temperature dependence of $H_{irr}$ for $H \parallel c$ evaluated by three different methods (i.e., $\rho(H, T)$, $J_c$-$H$, and $F_p$-$H$) is summarised in fig. 2c. $H_{irr}$ in the temperature range $20 \leq T \leq 35$ K from $J_c$-$H$ follow well the $H_{irr}$-line expressed by eq. (1) with an exponent $k$=1.2, which is close to the theoretically predicted value of 4/3 for a glass-liquid transition [24, 25].

$$\mu_0 H_{irr} = 36.6 \left(1 - \frac{T}{T_{irr}}\right)^k \cdots (1)$$

Here, $T_{irr}$ is the irreversibility temperature for self-field, which is 37.4 K. This result indicates that the criterion for determining $H_{irr}$ is quite reasonable and consistent. However, $H_{irr}$ starts to deviate from eq. (1) at around 15 K. A steep increase of $H_{irr}$ at low temperatures was also observed in LaFeAs(O,F) [26], where it was related to a similar increase of $H_{c2}$ at the same temperature. This is due to the 2-dimensional multiband character of the superconductivity of these compounds in contrast to the 3-dimensional multiband superconductor Co-doped $BaFe_2As_2$ [27], where such an increase of $H_{irr}$ and $H_{c2}$ was not observed.

**Angle dependence of $J_c$**

To further understand the pinning mechanism, the angular dependence of $J_c$ was measured at three different temperatures, $T$=10, 20, and 30 K (fig. 5). Simultaneously, the corresponding $n$ values in $E \sim J^n$ is also plotted. As expected from the microstructural observation, the minimum $J_c$ is always observed at $\theta$=0º (*i.e.*, $H \parallel c$), whereas the maximum $J_c$ is located at $\theta$=±90º (*i.e.*, $H \parallel ab$). Additionally, the $J_c$ peak at $H \parallel ab$ becomes sharper with increasing the applied field. Because the exponent $n$ is proportional to the pinning potential $U$, $J_c(T, H, \theta)$ should show a behaviour similar to $n(T, H, \theta)$ [28]. Indeed, this relation holds at 30 K. However, $n(\theta)$ at 20 K shows a dip at $\theta$ close to ±90º for applied magnetic fields exceeding 3 T. At an even lower temperature of 10 K, a peak located at the local minimum around $H \parallel ab$ is observed (see, fig. 5e: for clarity $n(\theta)$ at 14 T was plotted), which evolves with decreasing the field. Such behaviour can be explained by intrinsic pinning, as observed in *RE*BCO [28, 29, 30] and FBS [31, 32, 33], arising from the modulation of the superconducting order parameter along the crystallographic $c$-axis. Vortices depin from intrinsic pinning through the double-kink mechanism [34], which easily creep along the $ab$-plane, resulting in small $n$. Here, the flux creep rate is proportional to the inverse of $n$-1 [35]. The cross-over temperature $T_{cr}$ from 3-dimensional Abrikosov to 2-dimensional Josephson vortices is, accordingly, located between 20 and 30 K. To determine $T_{cr}$ precisely, $n(\theta)$ around $H$

|| $ab$ at 10 T with a step size of 1 K and $n(T)$ for $H \parallel ab$ under magnetic fields $5 \leq \mu_0 H \leq 14$ T were measured (Supplementary information, Figs. S3 and S4). As a result, $T_{cr}$ is determined as 24.5±0.5 K. Given that the FeAs layer spacing $d$ is 0.8437 nm determined by XRD, the out-of-plane coherence length at zero kelvin, $\xi_c(0)$, can be estimated by $\xi_c(0) = d\sqrt{\left(1-\frac{T_{cr}}{T_c}\right)/2}$ [21]. The resultant $\xi_c(0)$ is 0.39±0.01 nm, which is comparable to NdFeAs(O,F) [14, 33].

To decouple the pinning contributions arising from uncorrelated and correlated defects, the anisotropic Ginzburg-Landau (AGL) scaling [36] for the angle dependence of $J_c$ can be applied. This approach has been widely used for *RE*BCO [37] and FBS [26, 32, 33, 38]. In the absence of correlated pinning centres (*i.e.*, mainly randomly distributed and sufficiently small, isotropic pinning centres determine the pinning behaviour), all $J_c(\theta)$ curves at a given temperature collapse onto a single curve if plotted as a function of effective field $H_{eff}$:

$$H_{eff} = H\sqrt{\cos^2\theta + \frac{\sin^2\theta}{\gamma_J^2}}$$

where $\gamma_J$ is the anisotropy parameter. The AGL scaling, fig. 6, shows that some portion of $J_c(\theta)$ curves at given temperatures indeed scale with $H_{eff}$ when $\gamma_J$ is appropriately chosen. $\gamma_J$ decreases from 1.6 to 1.25 with decreasing temperature in contrast to NdFeAs(O,F) [14,33], where it increased. Clear deviations from the master curves due to the *ab* correlated pinning (here mostly intrinsic pinning because of the layered crystal structure) become obvious with decreasing temperature and also increasing field.

**Discussions**

Our NdFeAs(O,H) film shows a high self-field $J_c$ exceeding 10 MA/cm$^2$ at 5 K, which is a record level value for pnictides without artificial pinning centres. According to Talantsev and Tallon [39], self-field $J_c$ for type-II superconductors can be expressed by $H_{c1}/\lambda$, if the sample thickness is less than $\lambda$. Here, $H_{c1}$ is the lower critical field and $\lambda$ the relevant London penetration depth. Hence, the high self-field $J_c$ of NdFeAs(O,H) may be due to a short London penetration depth at heavily electron doping.

Another effect of heavily electron doping is the reduction of anisotropy. The $H_{c2}$ anisotropy near $T_c$ for NdFeAs(O,H) is $\gamma_{Hc2}$=4.4, which is smaller than that of NdFeAs(O,F) ($\gamma_{Hc2}$=5.1). Additionally, compared with NdFeAs(O,F), the temperature dependence of the anisotropy $\gamma_J$ evaluated from the AGL scaling for NdFeAs(O,H) [14, 33] shows an opposite behaviour. It is also worth mentioning that $\gamma_J$ of NdFeAs(O,H) is comparable to that of Co-doped BaFe$_2$As$_2$ [38].

Heavily electron doping by means of hydrogen substitution for oxygen in *Ln*FeAsO is a novel method to tune superconducting properties, whilst $T_c$ is maintained around 45 K, comparable to NdFeAs(O,F). For most FBS in contrast, a high carrier concentration reduces $T_c$. Additionally, this method is rather simple, once the parent $Ln$FeAsO films are fabricated. Now the parent compound can be fabricated by both pulsed laser deposition [40, 41] and molecular beam epitaxy (MBE) [42, 43]. Hence, our study motivates coated conductor preparation, for which films with thicknesses in the micrometer range are needed. However, a homogeneous H substitution for oxygen seems to be difficult in such thick films. Indeed, the H concentration showed to be inhomogeneous for 90-nm thick SmFeAs(O,H) films [11]. To realise $Ln$FeAs(O,H) coated conductors and eventually applications of hydrogen-doped $Ln$FeAsO, new approaches to a homogeneous H substitution should be explored.

To conclude, we have grown hydrogen-doped NdFeAsO epitaxial thin films. TEM investigations supported that hydrogen is homogenously distributed. Detailed electric transport measurements revealed the benefits of heavily electron doping to $Ln$FeAsO in terms of high self-field $J_c$ and low anisotropy without compromising $T_c$.

**Methods**
**Thin film fabrication**
Parent NdFeAsO was grown on MgO(001) at 800°C by MBE [40]. The structural characterisation by x-ray diffraction (XRD) confirmed that the 24-nm thick film was phase pure and epitaxially grown with (001)[100]NdFeAsO ∥ (001)[100]MgO. After structural characterisation by XRD, the NdFeAsO films were cut into pieces of approximately 5×10 mm$^2$ and subsequently sealed in an evacuated silica-glass tube filled with ~0.5 g of CaH$_2$ powder that serves as a hydrogen source. Here, it is important that the film surface is in direct contact with the CaH$_2$ powders to promote a topotactic chemical reaction. The sealed silica-glass tube was heated to 490°C at a rate of 100°C/h, held at this temperature for 36 h, and then cooled to room temperature at a rate of 100°C/h. The NdFeAs(O,H) film was also phase pure after processing, indicating that the crystalline quality is not compromised.

**Microstructural analysis by TEM.**
The cross-sectional samples for TEM observation were fabricated by focused ion beam. Atomic-resolution observations were performed using a transmission electron microscope (Titan Cubed 60–300 G2, Thermo Fisher Scientific) which is equipped with a spherical aberration corrector (DCOR, CEOS GmbH) for the probe-forming lens system. The microscope was operated in the scanning TEM (STEM) mode at an accelerating voltage of 300 kV. The convergence semi-angle of the electron probe was set to 18 mrad. The typical probe diameter was less than 0.1 nm. An annular dark field (ADF) detector was positioned to detect scattered electrons of an angular range from 38 to 184 mrad. In order to measure the lattice parameters as accurately as possible, we employed a drift corrected frame integration available in Velox software (Thermo Fisher Scientific) to avoid image distortion due to sample drifting. The magnification of each image was calibrated by the lattice parameters of the MgO substrates.

**Electrical transport properties.**

For temperature ($T$)-, field ($H$)-, and direction ($\theta$)-dependence measurements of resistivity, $\rho(T, H, \theta)$, and critical current density, $J_c(T, H, \theta)$, the NdFeAs(O,H) film was photolithographically patterned and Ar-ion beam etched to fabricate a small bridge of 30 μm width and 1 mm length. The sample was mounted on a rotator with maximum Lorentz force configuration, where the direction of the bias current is always perpendicular to that of the applied field. The angle $\theta$ is measured from the crystallographic $c$-axis. The critical temperature $T_c$ was determined as the intersection between the fit to the normal state resistivity and the steepest slope of resistivity. By measuring $T_c$ at various fields, the upper critical field $H_{c2}$ versus $T$ diagram was obtained. The bias current for resistivity measurements was 10 μA, corresponding to a current density of $J_b \sim 1.4$ kA/cm$^2$. The irreversibility field $H_{irr}$ was evaluated from $\rho(T, H)$ and $J_c(T, H)$ data. For the former $H_{irr}$ is determined by the intersection between the $\rho(T, H)$ curves and the resistivity criterion $\rho_c = E_c/J_b \sim 7.2 \times 10^{-7}$ mΩcm, where $E_c$ (1 μV/cm) is the electric field criterion for determining $J_c$ (Supplementary information fig. S5). For the latter $H_{irr}$ was determined by the intersection between $J_c(T, H)$ curves and $J_b$. At $H_{irr}$, the electric field - current density $J$ characteristics showed a relation that can be expressed as $E \sim J^n$, where $n$ was close to 1.

**Acknowledgements**

This work was supported by JST CREST Grant Number JPMJCR18J4, JSPS Grant-in-Aid for Scientific Research (B) Grant Number 20H02681 and Japan-German Research Cooperative

Program between JSPS and DAAD, Grant number JPJSBP120203506. A part of the work was also supported by Advanced Characterization Platform of the Nanotechnology Platform Japan sponsored by the Ministry of Education, Culture, Sports, Science and Technology (MEXT), Japan.

**Author Contributions**
K.I. and J.H. designed the study and wrote the manuscript together with H.S., S.H., and H.I. Thin film preparation, structural characterisations by XRD, and micro bridge fabrications were carried out by K.I., K.K., M.C., and T.H. Microstructural characterisations by TEM were performed by C.W., H.S. and S.H., and J.H. conducted electrical transport measurements.

**Figures Captions**

**Figure 1. Microstructural analysis by TEM. a.** TEM cross-sectional view of the NdFeAs(O,H) epitaxial thin film revealed almost no apparent defects. Additionally, no reaction layer between the film and the MgO substrate was observed. **b.** The magnified ADF images of NdFeAsO and **c.** NdFeAs(O,H). **d.** Image intensity profiles along the $c$-axis direction extracted from **b.** and **c.**, averaged in the $a$-axis direction. The $c$-axis lattice parameters averaged over 10 layers are 8.64 Å and 8.50 Å, respectively. The distances from the 1$^{st}$ to the 11$^{th}$ layer and from the 18$^{th}$ to the 28$^{th}$ layer are the same in the NdFeAs(O,H) film.

**Figure 2. Temperature dependence of the resistivity at various magnetic fields for the NdFeAs(O,H) epitaxial thin film.** Direction of the applied fields $H$ parallel to **a.** the $ab$-plane and **b.** the $c$-axis. **c.** Magnetic phase diagram. Irreversibility fields determined from the field dependence of $J_c$ and the corresponding $F_p$ for $H \parallel c$ follow well the irreversibility field line down to 20 K, which is expressed by eq. (1).

**Figure 3. Arrhenius plot of the field dependence of the resistivity traces shown in fig. 2.** Applied field parallel to **a.** the $ab$-plane and **b.** $c$-axis. **c.** Field dependence of the activation energy $U_0$ of thermally assisted flux motion for both main crystallographic directions. The exponent $\alpha$ for $H \parallel ab$ changes from 0.07 to 1 between 2 and 4 T, as indicated by a red arrow, before again changing to 0.34.

**Figure 4. In-field electrical transport $J_c$ characteristics and pinning force density $F_p$.** Field dependence of $J_c$ at various temperatures for **a.** $H \parallel ab$ and **b.** $H \parallel c$. Corresponding pinning force density $F_p$ for **c.** $H \parallel ab$ and **d.** $H \parallel c$. For comparison, the data of NdFeAs(O,F) measured at 4.2 K are superimposed [14]. **e.** Temperature dependence of exponents $p$ and $q$ in $f_p \sim h^p(1-h)^q$ for $H \parallel c$.

**Figure 5. Angular dependence of $J_c$ and the corresponding exponent $n$.** Measurement temperatures were **a.** 10 K, **b.** 20 K, and **c.** 30 K. Enlarged view of $n(\theta)$ in the vicinity of -90° is shown at the bottom of each panel. **d.** $n(\theta)$ at 10 and 20 K under a fixed field of $\mu_0H$=14 T. **e.** $n(\theta)$ at 10, 20 and 30 K under a fixed field of $\mu_0H$=5 T.

**Figure 6. Scaling behaviour of $J_c(\theta)$ as a function of effective field.** All $J_c(\theta)$ data except for those where the contribution of the *ab*-correlated pinning is dominant fall onto the measured curves of $J_c$ (i.e., field dependence of $J_c$ for $H \parallel c$ (lines), shown in fig. 4a) with $\gamma_J$ values of 1.25-1.6.

**Figure 1**

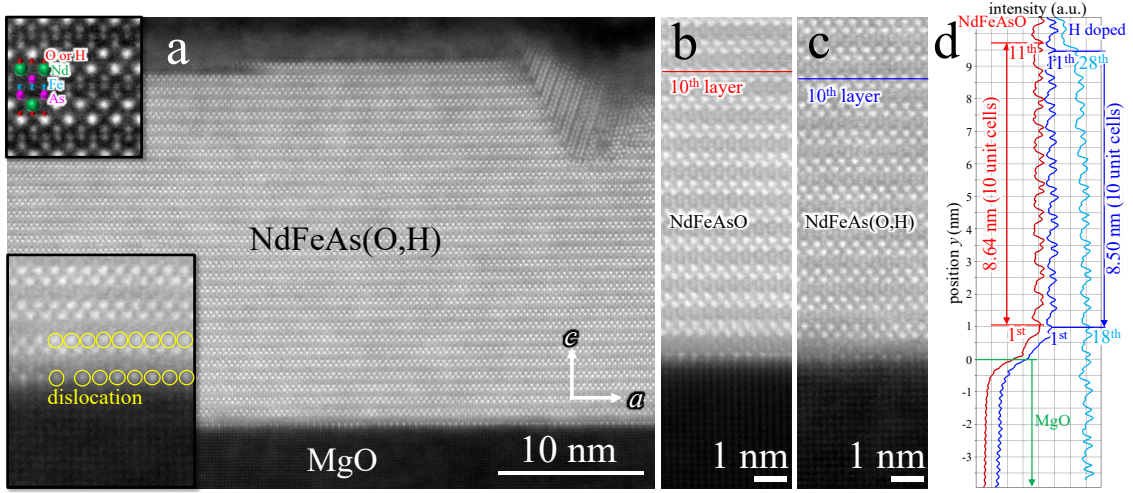

**Figure 2**

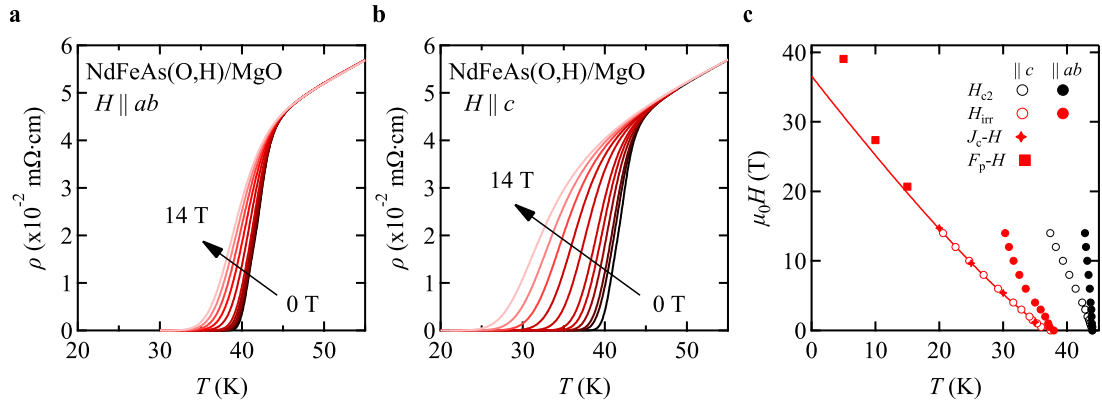

**Figure 3**

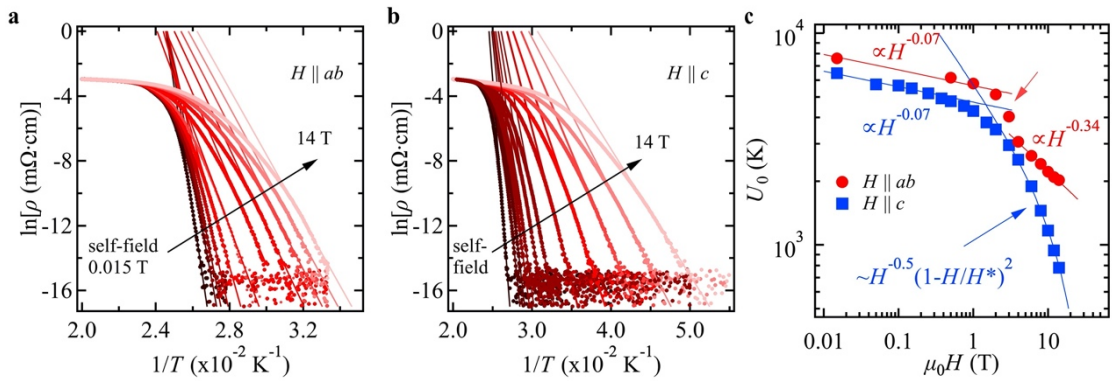

**Figure 4**

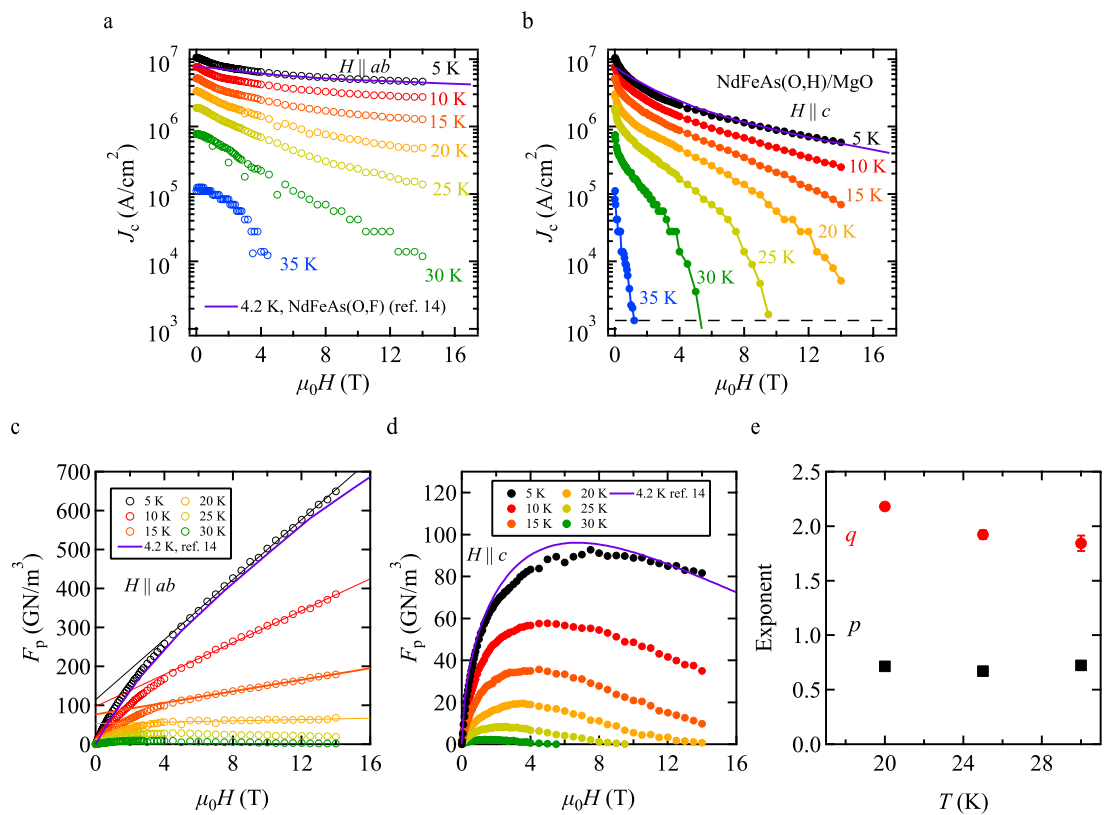

**Figure 5**

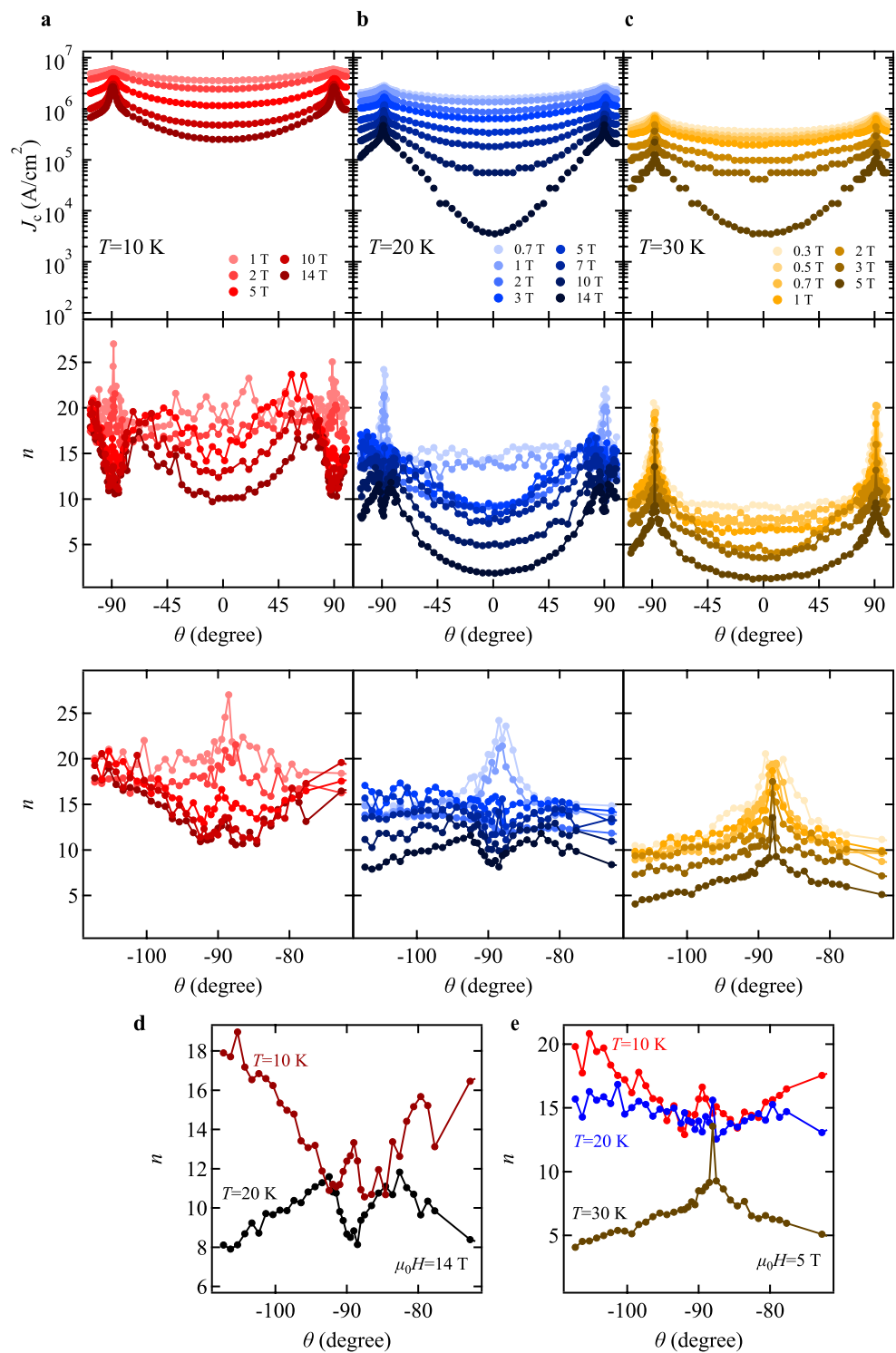

**Figure 6**

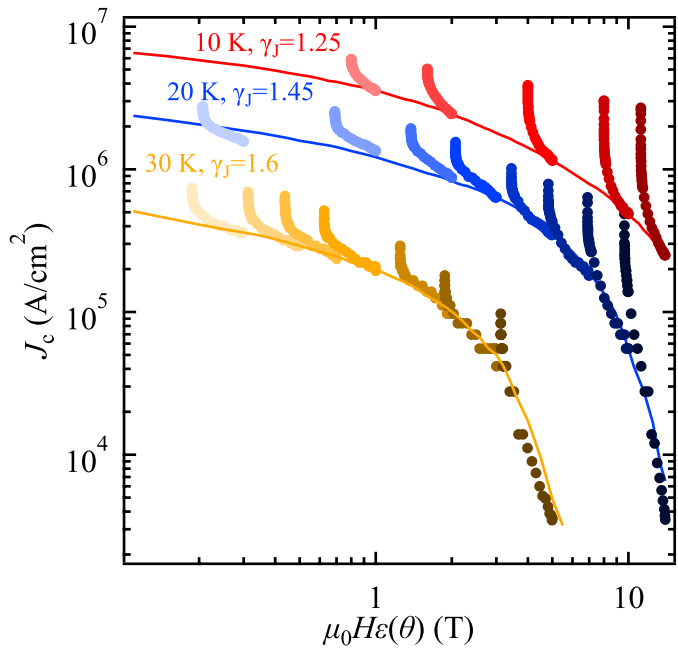

**Supplementary information on "High $J_c$ and low anisotropy of hydrogen doped NdFeAsO superconducting thin film"**

Kazumasa Iida[1,6], Jens Hänisch[2], Keisuke Kondo[1], Mingyu Chen[1], Takafumi Hatano[1,6], Chao Wang[3], Hikaru Saito[4,6], Satoshi Hata[3,5,6], and Hiroshi Ikuta[1]

[1]*Department of Materials Physics, Nagoya University, Chikusa-ku, Nagoya 464-8603, Japan*
[2]*Institute for Technical Physics, Karlsruhe Institute of Technology, Hermann-von-Helmholtz-Platz 1, 76344 Eggenstein-Leopoldshafen, Germany*
[3] *The Ultramicroscopy Research Center, Kyushu University, Nishi-ku, Fukuoka 819-0395, Japan*
[4]*Institute for Materials Chemistry and Engineering, Kyushu University, Kasuga, Fukuoka 816-8580, Japan*
[5]*Faculty of Engineering Sciences, Kyushu University, Kasuga, Fukuoka 816-8580, Japan*
[6]*JST CREST, Kawaguchi, Saitama 332-0012, Japan*

1. **The effect of microbridge processing on the superconducting transition temperature.**

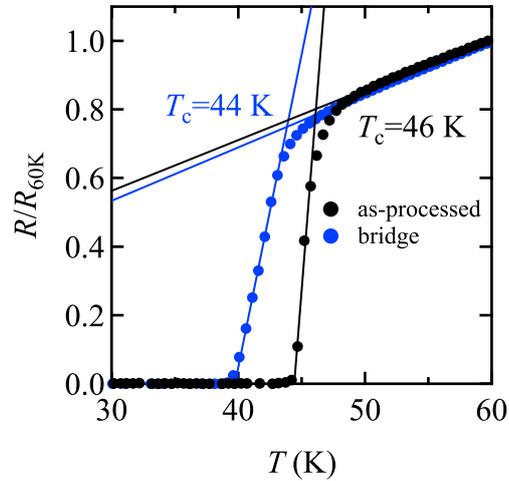

**Figure S1. Normalised resistance as a function of temperature.** The as-processed sample showed a $T_c$ of 46 K. After fabrication, the bridge sample showed a $T_c$ of 44 K.

2. **The pinning force analysis.**

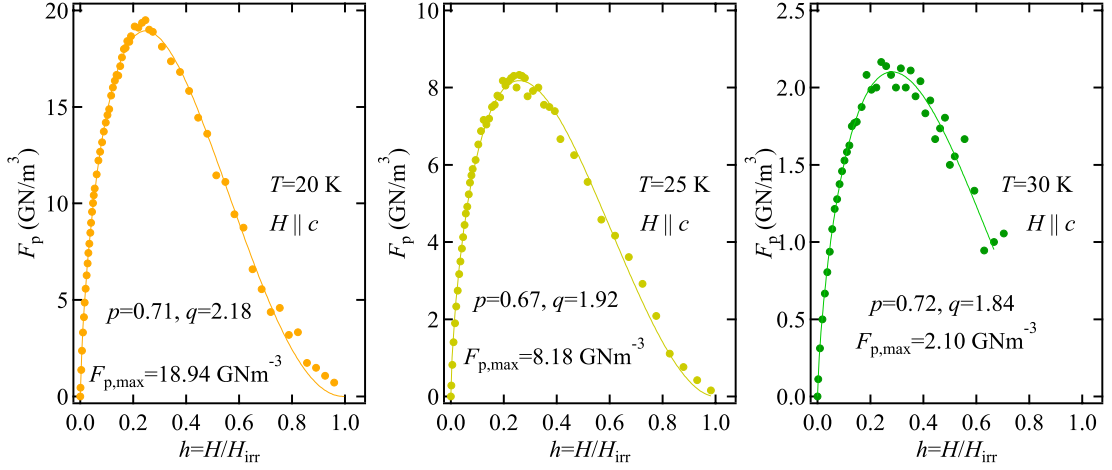

**Figure S2. Pinning force density $F_p$ as a function of reduced field for $H \parallel c$.** Exponents $p$ and $q$ in $h^p(1-h)^q$ are evaluated for each temperature. The results are summarised in fig. 4e.

3. **Determining the dimensional cross-over temperature $T_{cr}$.**

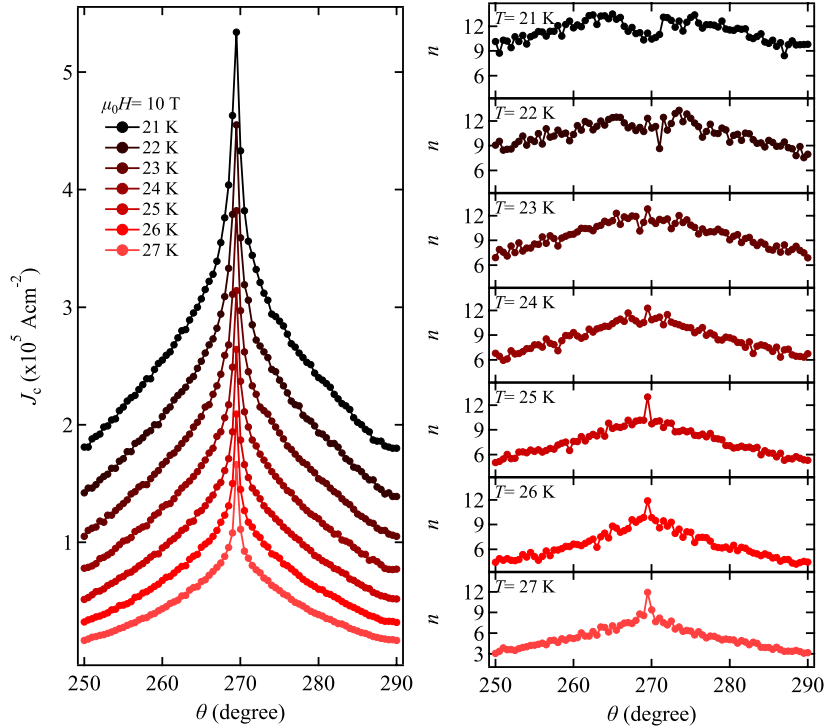

**Figure S3. Angular dependence of $J_c$ and the corresponding $n$ under a fixed magnetic field of 10 T measured at various temperatures.** As the temperature decreases from 27 K, the exponent $n$ around $\theta \sim 270°$ (i.e. $H \parallel ab$) starts showing shoulders at 24 K, followed by a dip formation.

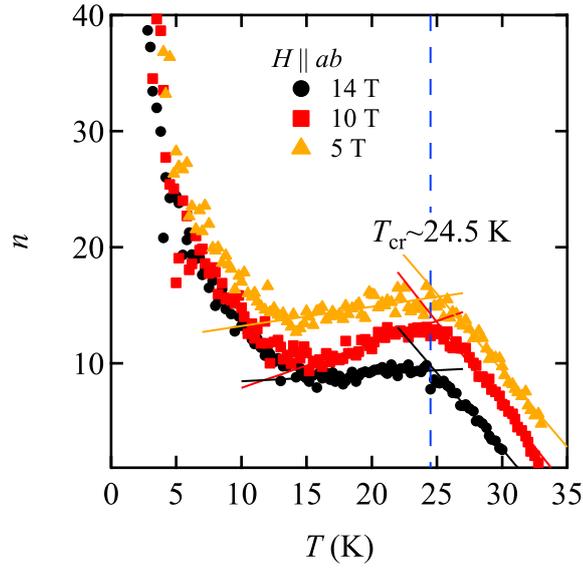

**Figure S4. Temperature dependence of *n* for *H* ∥ *ab* under several magnetic fields.** The exponent *n* increases with decreasing *T* and stays constant around 24.5 K. Below 15 K, *n* starts to increase again with decreasing *T*.

4. *E-J* curves for determining $J_c$.

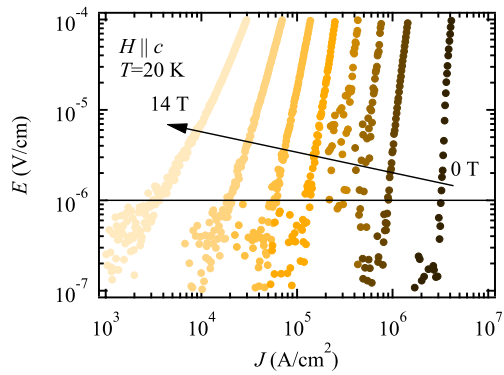

**Figure S5. Representative *E-J* curves.** *E-J* curves at 20 K for *H* ∥ *c*. Field increment was 2 T. $J_c$ was determined as the intersection between *E*=1 μV/cm and each curve.